\documentclass[aps,twocolumn,superscriptaddress,showpacs,floatfix]{revtex4}

\newcommand{\eps}{\epsilon}
\newcommand{\epsi}{\eps_{\infty}}
\newcommand{\epss}{\eps_{s}}
\newcommand{\wfx}{\psi(x)}
\newcommand{\wfix}{\psi_{i}(x)}
\newcommand{\norm}[1]{| #1 |^{2}}
\newcommand{\aB}{a_{B}}
\newcommand{\est}{\epsilon^{*}}
\newcommand{\Ry}{\mathrm{Ry}}
\newcommand{\kB}{k_{B}}
\newcommand{\Ebi}{E_b}

\newcommand{\Ebr}[1]{E^0_{#1}}

\newcommand{\Etot}[1]{\mathcal{E}_{#1}}
\newcommand{\Eto}{\mathcal{E}}
\newcommand{\Eon}{\Etot{1}}
\newcommand{\Etw}{\Etot{2}}
\newcommand{\Eron}{\Etot{\mathrm{r}1}}
\newcommand{\Ertw}{\Etot{\mathrm{r}2}}
\newcommand{\Etor}{\Etot{\mathrm{r}}}
\newcommand{\Ea}{\Eron}
\newcommand{\Eb}{\Ertw}
\newcommand{\Etr}{\Etot{\mathrm{tr}}}

\newcommand{\Eact}{E_{\mathrm{act}}}
\newcommand{\Er}{E_{\mathrm{r}}}

\newcommand{\Vknl}[1]{V_{#1}}

\newcommand{\xvec}{\mathbf{P}}
\newcommand{\Pa}{\mathscr{P}_1}
\newcommand{\Dp}{\mathscr{D}}
\newcommand{\tab}{t_{21}}
\newcommand{\rs}{r}
\newcommand{\rso}{\rs_1}
\newcommand{\rst}{\rs_2}
\newcommand{\rsot}{\rs_{12}}
\newcommand{\Rt}{R}
\newcommand{\rhx}{\rho(x)}
\newcommand{\rhix}{\rho_i(x)}

\newcommand{\secder}{\frac{\partial^2}{\partial x^2}}

\newcommand{\rf}{\rho^{f}}
\newcommand{\rh}{\rho}
\newcommand{\rfx}{\rho^{f}(x)}
\newcommand{\rfy}{\rho^{f}(y)}
\newcommand{\Hx}{\hat{H} (x)}
\newcommand{\Hix}{\hat{H}_i (x)}
\newcommand{\al}{\alpha}
\newcommand{\be}{\beta}
\newcommand{\wfcx}{\psi^{*} (x)}
\newcommand{\Vf}{\widetilde{V}}
\newcommand{\vrr}{\mathbf{r}}
\newcommand{\Ut}{\widetilde{U}}
\newcommand{\PE}{\mathscr{V}}
\newcommand{\PEi}{\mathscr{V}_i}
\newcommand{\tEto}{\widetilde{\Eto}}
\newcommand{\HH}{\langle H \rangle}
\newcommand{\NN}{\mathcal{N}}
\newcommand{\rfix}{\rho^{f}_{i}(x)}
\newcommand{\rfjy}{\rho^{f}_{j}(y)}
\newcommand{\rfox}{\rho^{f}_{1}(x)}
\newcommand{\rftx}{\rho^{f}_{2}(x)}
\newcommand{\rhox}{\rho_{1}(x)}
\newcommand{\rhtx}{\rho_{2}(x)}
\newcommand{\kab}{k_{1\rightarrow 2}}
\newcommand{\kba}{k_{2\rightarrow 1}}
\newcommand{\kband}{k_{1\rightarrow \mathrm{band}}}

\usepackage{graphicx}
\usepackage{amsmath}
\usepackage{latexsym}
\usepackage{amsfonts}
\usepackage{amssymb}
\usepackage{mathrsfs}

\begin{document}

\title{Solvation-induced one-dimensional polarons and electron transfer}

\author{G.~L.~Ussery}
\affiliation{Department of Physics, The University of Texas at
Dallas, P. O. Box 830688, EC36, Richardson, Texas 75083, USA}
\author{Yu.~N.~Gartstein}
\affiliation{Department of Physics, The University of Texas at
Dallas, P. O. Box 830688, EC36, Richardson, Texas 75083, USA}

\begin{abstract}
When a one-dimensional (1D) semiconductor nanostructure is immersed in a sluggish polar solvent, fluctuations of the medium may result in the appearance of localized electronic levels inside the band gap. An excess charge carrier can occupy such a level and undergo self-localization into a large-radius adiabatic polaron surrounded by a self-consistent medium polarization pattern. Within an appropriately adapted framework of the Marcus theory, we explore the description and qualitative picture of thermally activated electron transfer involving solvation-induced polaronic-like states by considering transfer between small and 1D species as well as between two 1D species. Illustrative calculations are performed for tubular geometries with possible applications to carbon nanotube systems.
\end{abstract}

\pacs{31.70.Dk, 71.38.-k, 82.20.Yn, 82.45.Yz}

\maketitle

\section{Introduction}

\begin{figure}
\centering
\includegraphics[scale=0.6]{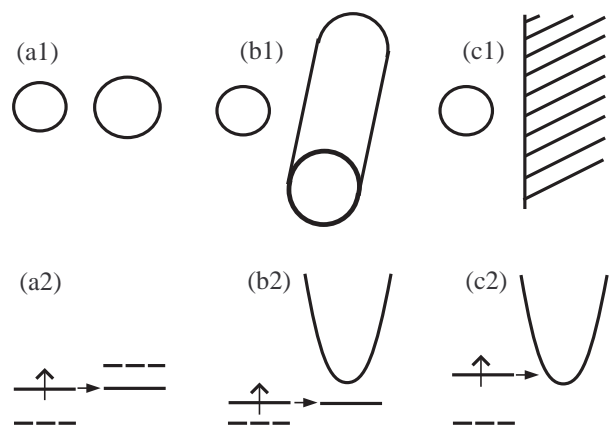}
\caption{\label{ctregimes} Schematic pictures of electron transfer (a) between two small species, (b) between a small and a 1D species, and (c) between a small species and a bulk semiconductor. The second row depicts electronic levels (and bands) and their modulation by medium fluctuations to promote iso-energetic electron transfer.}
\end{figure}

Electron transfer reactions mediated by fluctuations of the (polar) environment are of great importance for a variety of physico-chemical and biological processes. Descriptions of the Marcus theory of electron transfer and various elaborations on the topic are staples in numerous textbooks (e.~g., \cite{CTbook,CTbook1,CET2004,fawcett,nitzan}). In terms of traditional application areas, one may distinguish electron transfer between small species (chemical reactions in solutions), as schematically illustrated in Fig.~\ref{ctregimes}(a), and transfer between small species and bulk electrodes (electrochemistry), Fig.~\ref{ctregimes}(c). In the former case, the transfer occurs between two discrete electronic states; in the latter case the transfer is between a discrete level and a rigid continuous band of electronic states. The subject of our interest in this paper is thermally activated electron transfer to/from one-dimensional (1D) semiconducting nanostructures, as sketched in Fig.~\ref{ctregimes}(b) and exemplified by systems like conjugated polymers \cite{Barford_book}, nanotubes and nanowires \cite{SCelectrodes}. 1D systems are known to be particularly sensitive to potential energy fluctuations resulting in the formation of bound states below the continuum even in shallow potential wells \cite{LL3}. In fact, when an excess charge carrier is added to a 1D semiconductor, the long-range Coulomb interaction with the surrounding polar solvent may result in self-trapping of the carrier, where a localized electronic state is accompanied by a self-consistent dielectric polarization pattern of the medium. Formation of such solvation-induced 1D polarons has been discussed recently both within simplified theoretical models  \cite{basko,YNGpol, polcylinder,GU_lowfreq,UG_optabs} and at the level of \textit{ab initio} computations \cite{MG_abinitio1,MG_abinitio2}. The sketch of electron transfer in Fig.~\ref{ctregimes}(b) involving polaron-like localized electronic states is thus, in a sense, intermediate between traditional pictures of Fig.~\ref{ctregimes}(a) and (c).

It should be noted here that 1D semiconductor nanostructures in contact with polar solvents are candidates for various technological applications involving fundamental redox reactions, such as in photoelectrochemistry, energy harvesting and sensing (e.~g., \cite{SCelectrodes,pyang05,photcat2,kamat07,csc08,mattod09,yurev,sens1}). As we will be using later specific illustrative calculations for small-diameter tubular structures, semiconducting single-wall carbon nanotubes (SWCNTs) are a particularly widely known example of such. Redox chemistry of CNTs has been deemed an ``emerging field of nanoscience'' \cite{chirsel} and solvatochromic effects in CNTs are being intensely researched \cite{CNTsolvchrom}.

Considerations of the electron transfer problem \cite{CTbook,CTbook1,CET2004,fawcett,nitzan} involve various aspects and regimes. In this paper we are concerned only with a very limited scope of questions. The regime we are interested here is usually referred to as  high-temperature ($T$), non-adiabatic transfer epitomized by the famed Marcus expression (e.g., \cite{CET2004,nitzan})
\begin{equation}\label{clMarcus}
\kab=\frac{2 \pi}{\hbar}\frac{\norm{\tab}}{\sqrt{4\pi \Er \kB T}}\exp\left(-\frac{\Eact}{\kB T}\right)
\end{equation}
for the transfer rate between discrete electronic states $1$ (``donor'') and $2$ (``acceptor''). Here $t_{21}$ is the electronic transition matrix element, while the activation energy
\begin{equation}\label{acten}
\Eact=\frac{(\Eb-\Ea+\Er)^2}{4 \Er}
\end{equation}
is determined by the difference of the fully equilibrated (free) energies $\Eb$ and $\Ea$ of the respective states, and the so-called reorganization energy $\Er$.

We restrict our consideration to the outer sphere reorganization referring to the interaction of the charge carrier with the polarization $\xvec (\vrr)$ of the polar medium. (For a related discussion of the effects of the interaction with 1D lattice distortions in the context of the electron injection into a deformable polymer chain, see Ref.~\cite{Basko_injection}.) Expression (\ref{clMarcus}) can be derived from the averaging of the Fermi golden rule
\begin{equation}\label{FermiGR1}
\kab =\frac{2 \pi}{\hbar} \int \Dp\xvec\, \Pa (\xvec)\, \norm{\tab(\xvec)}\, \delta(E_1(\xvec)-E_2(\xvec))
\end{equation}
for the electronic transition over various spatial distributions ($\Dp\xvec $) of harmonic solvent fluctuations under assumptions of the Boltzmann probability $\Pa (\xvec)$ with the electron in state $1$, linear coupling of purely electronic energies $E_1$ and $E_2$ to fluctuations, and $\tab$ being independent of $\xvec$. Equation (\ref{FermiGR1}) very nicely reflects the spirit of the Marcus theory, where iso-energetic electron transfer takes place in configurations created by appropriate solvent fluctuations. In what follows we discuss applications of this approach to electron transfer between a small species and a 1D semiconductor as well as between two parallel 1D structures. Our emphasis would be on the qualitative picture of \textit{thermally activated} transfer and on the evaluation of the corresponding activation energies.

While, for certainty, we explicitly discuss the transfer of an electron in the context of the conduction band of a 1D semiconductor, it should be absolutely clear that the same picture applies to the transfer of a hole in the context of the valence band of the semiconductor. One could as well use the language of the electronic excitation energies equally applicable to both electrons  and holes. Also, the principle of detailed balance in thermal equilibrium allows one to immediately relate rates $\kab$ of the forward (``to'') and $\kba$ of the reverse (``from'') processes \cite{CET2004,nitzan} and their respective activation energies.

\section{\label{Solv1}Electron solvation on small species and on 1D semiconductor}

As an introduction to the further discussion, in this section we compare solvation of an electron (charge $q$) separately on a small spherical species  and on a 1D semiconductor structure. The question of interest here is a relationship between the (lowest) purely electronic energy $E$ as affected by the solvent polarization $\xvec (\vrr)$ and the \textit{minimal possible} total free energy $\Eto$ of the electron-solvent system for a given $E$. The total energy includes, besides $E$, the (free) energy  $U$ stored in the solvent polarization. Our question of interest can also be expressed as that of \textit{minimal possible} energy $U$ for a given $E$. The bare, unaffected by the polarization, electronic energy will be denoted as $\Ebr{}$.

We find it very convenient to represent the state of solvent polarization $\xvec (\vrr)$ by a distribution of fictitious charges $z_i$ that would cause this polarization pattern as an equilibrium dielectric response \cite{nitzan,YNGpol}. In terms of such charges, the energy stored in the polarization would then be generically given as
\begin{equation}\label{genPolE}
U=\frac{1}{2}\sum_{ij}\Vknl{ij}z_i z_j,
\end{equation}
while the contribution to the electronic potential energy described by
\begin{equation}\label{genEl}
-q\sum_{i}\Vknl{ei}z_i.
\end{equation}
The interaction kernel elements $\Vknl{ij}$ and $\Vknl{ei}$ are determined by solutions of the corresponding electrostatic potential problem for specific spatial distributions (and positions) of fictitious charges $z_i$ and the real charge $q$.

One recalls that polarization $\xvec$ relevant for the electron transfer problem is not the total solvent polarization but its slower (orientational) component \cite{CTbook,CTbook1,CET2004,fawcett,nitzan}. For a polar environment, such as the Debye solvent, characterized by two dielectric constants: static $\epss$ and high-frequency $\epsi$, the effective dielectric constant $\est$ is usually defined via
$$ 
1/\est = 1/\epsi - 1/\epss
$$ 
to represent the dielectric response due to the slower polarization component. (For typical \cite{fawcett,YNGpol} solvents,  $\epss \gg \epsi$ so that $\est \simeq \epsi$.) The fast component of the total polarization acts ``instantaneously'' in the present context, and its effect is considered included in the definition of the bare electronic energies $\Ebr{}$. For simplicity, we restrict our attention to the uniform dielectric environment, see Refs.~\cite{YNGpol,UG_optabs} for a discussion of the effects of various dielectric conditions.

\begin{figure}
\centering
\includegraphics[scale=0.55]{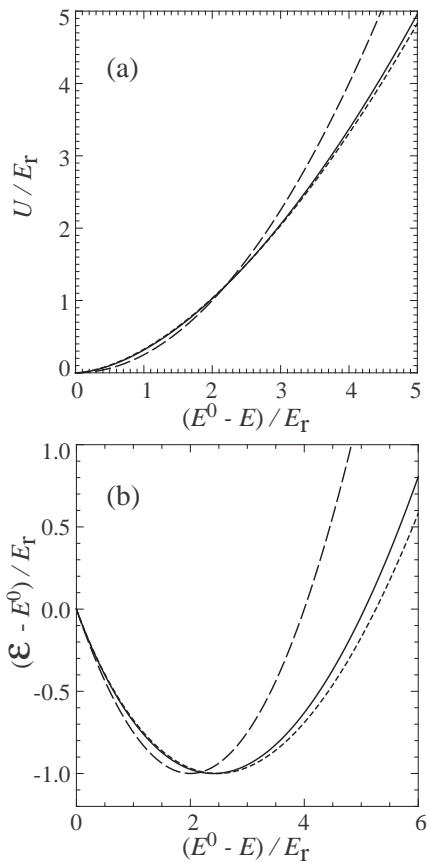}
\caption{\label{UETEE} Scaled functional dependences of the optimal (a) polarization energy $U$ and (b) total energy $\Eto$ for a given purely electronic energy $E$. Equlibrated solvation corresponds to the minima of total energies in panel (b). Long-dash lines show the benchmark parabolic behavior for small species; solid lines describe solvation on a 1D species with the confinement parameter $\aB/R=4$, short dashes are for the increased confinement of $\aB/R=10$.}
\end{figure}

The well-known case of a small spherical species of radius $\rs$ is very simple as it corresponds to the fixed on-the-sphere distributions for both charge $q$ and single fictitious charge $z$. That results in  $\Vknl{11}=\Vknl{e1}=1/\est\rs$. The electronic energy in this case is given entirely by Eq.~(\ref{genEl}):
\begin{equation}\label{small1}
E \{z\} = \Ebr{} - qz/\est\rs,
\end{equation}
while the polarization energy (\ref{genPolE}) by
\begin{equation}\label{small2}
U \{z\} = z^2/2\est\rs.
\end{equation}
The total system energy $\Eto \{z\} = E\{z\}+U\{z\}$ is a familiar parabola achieving its minimum of $\Etor=\Ebr{}-\Er$
at $z=q$ with the individual reorganization energy
\begin{equation}\label{small3}
\Er = q^2/2\est\rs.
\end{equation}
As is evident from Eqs.~(\ref{small1})-(\ref{small3}), the question we posed in the beginning of this section is answered by the parabolic dependence
\begin{equation}\label{small4}
U (E) = (E-\Ebr{})^2/4\Er
\end{equation}
for the small-species case. These benchmark dependences of $U (E)$ and $\Eto (E)=E+U(E)$ are shown in Fig.~\ref{UETEE} in terms of the energy scale (\ref{small3}).

Description of the adiabatic electron solvation on a 1D semiconductor is more involved as both the real (electron) charge $q$ and the fictitious charge $z$ can have variable spatial extents along it (coordinate $x$):
\begin{equation}\label{dens1}
q=\int dx \, \rhx, \ \ \ \ z= \int dx \, \rfx,
\end{equation}
where $\rhx$ and $\rfx$ are the respective linear charge densities. The former is determined by the (normalized) electron wave function $\wfx$:
\begin{equation}\label{dens2}
\rhx=q\norm{\wfx},
\end{equation}
following from the solution of the Schr\"{o}dinger equation
\begin{equation}\label{SE}
\Hx\wfx = (E-\Ebr{})\wfx.
\end{equation}
The Hamiltonian
\begin{equation}\label{Ham1}
\Hx = -\frac{\hbar^2}{2m}\secder + \PE (x)
\end{equation}
here is composed of the kinetic energy with the effective mass $m$ and the potential energy due to the solvent polarization:
\begin{equation}\label{PE}
\PE (x)=-q\int dy\ V (x-y)\, \rfy.
\end{equation}
This potential energy (\ref{PE}) is of course just a continuum version of Eq.~(\ref{genEl}), while the corresponding energy of the polarization (\ref{genPolE}) is now given by
\begin{equation}\label{NTPolE}
U\{\rf\} = \frac{1}{2}\int dx\, dy\, \rfx\, V (x-y)\, \rfy.
\end{equation}
An explicit example of the kernel $V(x)$ is discussed below.

It is understood that only the lowest energy eigenvalue $E < \Ebr{}$ is of relevance for us here. We now want to find configurations with the minimal possible energy $U$ (\ref{NTPolE}) that would yield a given value of $E$. The constrains imposed are conveniently taken into account in the minimization of the auxiliary functional
\begin{equation}\label{FU1}
\Ut = U  + \al \left(\HH - (E-\Ebr{}) \right) + \be \NN,
\end{equation}
where $\al$ and $\be$ are the Lagrange multipliers,
$$
\HH = \int dx\,  \wfcx \Hx \wfx
$$
and
\begin{equation}\label{Norm}
\NN = \int dx\, \norm{\wfx} - 1.
\end{equation}
One immediately finds that the variation of $\Ut$ in Eq.~(\ref{FU1}) over $\wfcx$ would lead to Eq.~(\ref{SE}) for $\be = \al (\Ebr{}-E)$. The variation over $\rfx$, on the other hand, establishes that
\begin{equation}\label{Fic1}
\rfx = \al \rhx,
\end{equation}
with $\rhx$ defined by Eq.~(\ref{dens2}).
Comparison with Eq.~(\ref{dens1}) shows that multiplier $\al$ has the meaning of the relative magnitude of the fictitious charge: $\al = z/q$.

Using $\al$ parametrically, one can now solve a self-consistent problem of Eqs.~(\ref{SE}), (\ref{dens2}), (\ref{Fic1}), and (\ref{PE}) iteratively to find the required dependences of $U (E)$ and $\Eto (E) = E + U (E)$. The minimum of $\Eto = \Ebr{}-\Er$ is achieved at $\al=1$ -- this is the equilibrium polaron state. The individual reorganization energy $\Er$ in this case is called the polaron binding energy, $\Ebi$. We have already discussed at length \cite{YNGpol,UG_optabs} the spatial extent and the binding energy of equilibrium solvation-induced polarons; for a practical range of parameters, the \textit{scales} of those quantities are respectively established by   the Bohr radius and Rydberg energy for the corresponding 3D Coulomb problem:
\begin{equation}\label{defaB}
\aB=\est \hbar^2 / m q^2, \ \ \ \ \Ry=q^2/2\est\aB.
\end{equation}
For a perspective, if one were to take representative values of $\est=3$ and of the effective mass of $0.05\,m_e$ (that would correspond to SWCNT of radius $R \simeq 8$ {\AA} \cite{NTEffectMass}), then Eq.~(\ref{defaB}) would result in  $\aB \simeq 32$ {\AA} and $\Ry \simeq 76$ meV. A twice larger effective mass (say, in a smaller radius SWCNT \cite{NTEffectMass})  would decrease $\aB$ and increase $\Ry$ by a factor of 2; large variations can also be induced by changes in $\est$.

Figure \ref{UETEE} shows an example of numerical results for $U (E)$ and $\Eto (E)$ scaled with respect to polaron's $\Er=\Ebi$. The deviations from the parabolic dependences (\ref{small4}) of the small species are evident. As it frequently takes place for excitonic and polaronic problems in 1D \cite{pedersen,tersscaling,GU_lowfreq,UG_optabs}, a power-law turns out to be a good \textit{approximate} representation in this case as well:
$$
U/\Er \simeq 0.3 ((\Ebr{}-E)/\Er)^{1.7}.
$$
This approximate relationship is, of course, not universal and small parameter-dependent deviations are illustrated in the figure.
It should also be noticed that the standard relationship $\Er=(\Ebr{}-E)/2=U$ for equilibrated solvation on small species does not hold for a 1D polaron (compare minima of the curves in Fig.~\ref{UETEE}(b)). This is a consequence of the additional degrees of freedom (variable spatial extent) in the polaron case.

For the numerical calculations above and in the examples that follow, we use electrostatic kernels corresponding to tubular (nanotube) charge distributions of  transverse radius $\Rt$. As is well-known \cite{haugbook}, continuum models with the purely 1D Coulomb interaction lead to diverging excitonic and polaronic bindings. Taking into account the transverse distribution of the charge (rings in the case of the tubular geometry) regularizes that behavior, and the binding becomes a growing function of the confinement parameter $\aB/R$ (e.g., \cite{pedersen,YNGpol,UG_optabs}). For nanotubes, the kernel $V(x)$ in Eqs.~(\ref{PE}) and (\ref{NTPolE}) is conveniently expressed via its Fourier-transform \cite{ando,YNGpol}
$\Vf (k) = \int dx\, \exp(-ikx) \, V(x)$:
\begin{equation}\label{Kern1}
\Vf (k) = 2I_0(kR)K_0(kR)/\est,
\end{equation}
where $I$ and $K$ are the modified Bessel functions appearing in electrostatic problems with cylindrical symmetry \cite{jackson}. In earlier work \cite{polcylinder} we established that the 1D polaron picture is valid for $\aB/R \gtrsim 1$; for \textit{illustrative} calculations in this paper we use some representative values \cite{UG_optabs} of the confinement parameter. For specific applications to semiconducting SWCNTs, one should note that the effective mass in that case scales with tube's radius, approximately as $m \propto 1/R$ \cite{NTEffectMass}, which makes the confinement parameter nearly $R$-independent.

\section{\label{SSLS}Transfer between small species and 1D polaronic states}

In order to illuminate the commonalities and differences with the traditional Marcus setup, we start here with a brief exposition of the well-known case of transfer between two small spherical species. Our discussion is similar to the line of presentation used in Ref.~\cite{nitzan}.

Denoting fictitious charges on donor and acceptor species as $z_1$ and $z_2$, respectively, the modulation of electronic energies in states 1 and 2 in this case is as given by Eq.~(\ref{genEl}):
\begin{equation}\label{mar1}
E_1=\Ebr{1} - \frac{q}{\est}\left(\frac{z_1}{\rso} + \frac{z_2}{\rsot}\right),
\end{equation}
\begin{equation}\label{mar2}
E_2=\Ebr{2} - \frac{q}{\est}\left(\frac{z_2}{\rst} + \frac{z_1}{\rsot}\right),
\end{equation}
where $\rso$ and $\rst$ are the radii of the species and $\rsot$ is the distance between their centers.
The polarization energy (\ref{genPolE}) is then
\begin{equation}\label{mar3}
U  = \frac{1}{\est}\left(\frac{z_1^2}{2\rso}+\frac{z_2^2}{2\rst}+\frac{z_1 z_2}{\rsot}\right).
\end{equation}
The equilibrated solvation on each of the species corresponds to the unconstrained minimization of total energies $\Eto$ as discussed in Sec.~\ref{Solv1} with the same results. One immediately finds that minimization of $\Eon=E_1+U$ yields the minimum value of $\Eron=\Ebr{1}-q^2/2\est\rso$ at $z_1=q$, $z_2=0$. The minimization of $\Etw=E_2+U$, on the other hand, yields the minimum value of $\Ertw=\Ebr{2}-q^2/2\est\rst$ at  $z_1=0$, $z_2=q$.

While the Marcus result (Eqs.~(\ref{clMarcus}) and (\ref{acten})) directly follows from employing expressions (\ref{mar1})--(\ref{mar3}) in Eq.~(\ref{FermiGR1}), here we instead exclusively focus on the electron transition point. The optimal state of polarization at the transition is again determined by the minimum of the total energy but now constrained by the condition $E_1=E_2$ of equal purely electronic energies (which is the same as equal total energies: $\Eon=\Etw$). Minimization of the auxiliary energy function
\begin{equation}\label{mar4}
\tEto = U + E_1 + \al (E_2 - E_1)
\end{equation}
with the Lagrange multiplier $\al$ results in relationships that will be recurring in what follows:
\begin{equation}\label{part1}
z_1=(1-\al)q, \ \ \ z_2=\al q, \ \ \ z_1+z_2=q.
\end{equation}
The parameter $\al$ thus has a meaning of the partitioning of the charge $q$ between fictitious charges at the transition point. Optimization of Eq.~(\ref{mar4}) yields its value at the transition as
\begin{equation}\label{mar5}
\al = \left(\Ertw-\Eron+\Er \right)/2\Er,
\end{equation}
where reorganization energy
\begin{equation}\label{mar6}
\Er=\frac{q^2}{\est}\left(\frac{1}{2\rso}+\frac{1}{2\rst}-\frac{1}{\rsot} \right)
\end{equation}
consists of donor and acceptor individual reorganization energies corrected by the interaction term. As used in Eq.~(\ref{part1}), parameter $\al$ can also play a role of the convenient one-dimensional ``reaction coordinate'' \cite{nitzan}: from $\al=0$ for the electron in equilibrated state 1 to $\al=1$ for electron in state 2. The resulting parametric dependence of total energies in this case is that of famous Marcus parabolas:
\begin{equation}\label{mar7}
\Eon = \Eron + \Er \al^2, \ \ \ \ \Etw=\Ertw + \Er (1-\al)^2.
\end{equation}
The transition value of $\al$ in Eq.~(\ref{mar7}) at which $\Eon=\Etw=\Etr$ is of course given by Eq.~(\ref{mar5}). The activation energy of
the $1 \rightarrow 2$ transition:
\begin{equation}\label{mar8}
\Eact=\Etr-\Eron=\Er\al^2,
\end{equation}
is precisely the familiar Eq.~(\ref{acten}).

The modification of the above approach to the case of the acceptor being a 1D semiconductor species is done in the spirit of the development in Sec.~\ref{Solv1}, where the linear distributed charge densities (\ref{dens1}), (\ref{dens2}) have been used. Particularly, the fictitious charge on the acceptor now becomes
$$
z_2 = \int dx\, \rfx,
$$
while the energy of the electron on the donor becomes
\begin{equation}\label{new1}
E_1=\Ebr{1} - q z_1/\est\rs - q \int dx\, \Vknl{12} (x)\,\rfx,
\end{equation}
where $\rs$ is the donor radius, and the interaction kernel $\Vknl{12}$ is exemplified below. The analogy between (\ref{new1}) and earlier used (\ref{mar1}) is evident.

On the other hand, the energy of the polaronic-like electron state on the acceptor, $E_2-\Ebr{2}$ would now be determined by the lowest-energy solution of the Schr\"{o}dinger equation (\ref{SE}) with the potential
\begin{equation}\label{PE1}
\PE (x)= - q\int dy\ V (x-y)\, \rfy \  - q z_1 \Vknl{12} (x).
\end{equation}
We reiterate that the range of parameters we are interested in corresponds to the bound electronic states below the band edge: $E_2 < \Ebr{2}$. Equation (\ref{PE1}) differs from Eq.~(\ref{PE}) by its second term, which represents the interaction with the fictitious charge on the donor (in close analogy to a similar interaction term in Eq.~(\ref{mar2})).

Finally, the place of the polarization energy (\ref{mar3}) in the Marcus setup is now taken by
$$
U = U_1 \{z_1\} + U_2 \{\rf\} + z_1 \int dx\, \Vknl{12} (x)\, \rfx
$$
where $U_1\{z_1\}$ corresponds to the fictitious charge on small species (\ref{small2}), $U_2 \{\rf\}$ to the linearly distributed fictitious charge (\ref{NTPolE}), and the last term is their interaction. Using, for certainty, a tubular transverse distribution of 1D charges on a tube of radius $R$, the intratube interactions $V (x)$ are described by Eq.~(\ref{Kern1}) while the interaction $\Vknl{12} (x)$ with the small spherical species by the Fourier-transform
\begin{equation}\label{Kern2}
\Vf_{12} (k) = 2I_0(kR)K_0(kd)/\est,
\end{equation}
where $d$ is the distance between the tube axis and the center of the sphere.

Unconstrained (in electronic energy) minimization of total energies $E_1+U$ and $E_2+U$ here results, of course, in equilibrated solvation of small species (at $z_1=q$, $z_2=0$) and 1D polaron (at $z_1=0$, $z_2=q$), respectively, as described in Sec.~\ref{Solv1}. The optimal polarization fluctuation for iso-energetic electron transfer is constrained by the condition $E_1=E_2$ and found by minimization of the auxiliary functional
\begin{equation}\label{new3}
\tEto = U + E_1 + \al (E_2 - E_1) + \be \NN,
\end{equation}
where Lagrange multipliers $\al$ and $\be$ are immediately recognized by comparing to Eqs.~(\ref{FU1}) and (\ref{mar4}). It then comes as no surprise that variations of Eq.~(\ref{new3}) over $\wfcx$ and $\rfx$ lead to earlier found results:
$
\be = \al (\Ebr{2}-E_2)
$
for normalization (\ref{Norm}) related coefficient, and Eq.~(\ref{Fic1}) for the relationship between fictitious and real charge densities. Furthermore, variation over $z_1$ results in $z_1=(1-\al)q$, and, hence, the same type (\ref{part1}) of charge $q$ partitioning between fictitious charges is recovered at the transition point that was encountered in the conventional Marcus problem. Evaluating then the difference between the total energy $\Etr$ at the optimal fluctuation and the equilibrated energy $\Eron$, one arrives at the activation energy for the $1 \rightarrow 2$ transition as
\begin{equation}\label{new4}
\Eact =\left( \frac{q^2}{2\est\rs} + U_2\{\rh\} - q\int dx\, \Vknl{12} (x)\, \rhx \right)  \al^2.
\end{equation}

While the structure of Eq.~(\ref{new4}) is analogous to the structure observed in Eqs.~(\ref{mar8}), (\ref{mar6}), there is a subtle difference between them in terms of the dependence on the value of the partitioning parameter $\al$ at the transition point. The parameter $\Er$ in Eq.~(\ref{mar8}) is independent of $\al$, and therefore the $\al$-dependence of the activation energy there is strictly parabolic. Differently, the charge density $\rhx$ featured in Eq.~(\ref{new4}) is \textit{not} the equilibrated polaron density (which would be obtained for $\al=1$) but, rather, the self-consistently determined transition state density generally dependent on $\al$ itself (determined, in turn, by the equilibrated
energy offset between the species). Thus, rigorously speaking, the $\al$-dependence in Eq.~(\ref{new4}) is not strictly parabolic. By the same token, if one were to consider the $2 \rightarrow 1$ transition, there would be no strict symmetry between forward and reverse transitions as seen in Eq.~(\ref{mar7}). These subtle features are related to a qualitative difference in the details of solvation on small and 1D species (Sec.~\ref{Solv1}) and reflect the presence of new degrees of freedom for the solvated electron in variability of its spatial extent on the 1D species.

It is another matter that, quantitatively, the effect due to the small species can easily dominate, largely disguising those qualitative differences and deviations from the conventional parabolic behavior. Also contributing to this ``disguise'' is a compensatory character of the effects related to the second and third terms in parentheses of Eq.~(\ref{new4}). The former originates in electrostatic interactions within the 1D system, while the latter in interactions between the small and 1D species.

\begin{figure}
\centering
\includegraphics[scale=0.6]{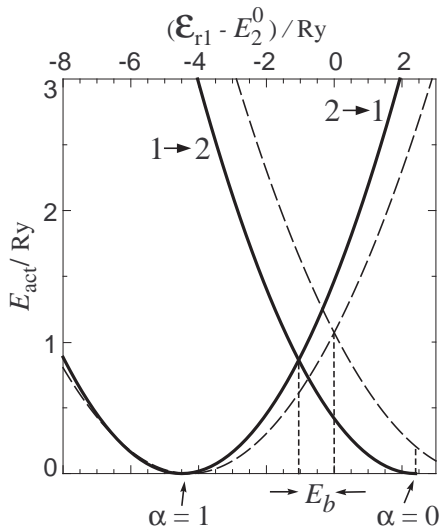}
\caption{\label{SSLSEact} Example of calculated activation energies for electron transfer between a small (state 1) and a 1D (state 2) species  as a function of the energy offset between the equilibrium energy of the small species and the band edge of the 1D semiconductor. The dashed lines show the results one would obtain for a rigid electronic band unaffected by the solvent polarization. See text for details.}
\end{figure}

Figure \ref{SSLSEact} depicts results of \textit{illustrative} calculations of the activation energies for the case characterized by the geometric parameters $r=3.5$ \AA, $R=5$ \AA, $d=10$ \AA, and $\aB=15$ \AA \ in the corresponding $\Ry$ units (\ref{defaB}). Shown are activation energies for both forward $1 \rightarrow 2$ and reverse $2 \rightarrow 1$ transitions. The results calculated as described above (solid lines) are compared to what one would get if the band states were unaffected by the solvent polarization (a rigid band, long-dash lines). In the latter case, the reorganization energy $\Er$ would be entirely due to the small species, and the activation energy for the $1 \rightarrow 2$ transition is well represented by the approximation $\Eact=(\Ebr{2}-\Ea+\Er)^2/4\Er$ (as one can numerically confirm from the conventional \cite{nitzan} integration of Eq.~(\ref{clMarcus}) over band states:
$$
\kband= \int_{\Ebr{2}}^{\infty} dE_2\, D(E_2)\, \kab (E_2),
$$
with 1D density-of-states $D(E_2)\propto 1/\sqrt{E_2-\Ebr{2}}$.) The relative displacement of the ``center lines'' shown in the figure indicates the equilibrium polaron binding energy $\Ebi$ (which is close to 1 $\Ry$ in this case).

Marked in Fig.~\ref{SSLSEact} with ``$\al=0$'' and ``$\al=1$'' are energy offsets where the respective activation energies vanish. It is worthwhile to stress here that the $\al=0$ point corresponds to the equilibrium polarization pattern of the small species (no fictitious charge on the acceptor (\ref{part1})). Still, the transition occurs into an electronic level below the band edge as is clearly seen by comparison with the dashed-line rigid-band result. This localized state on the acceptor is entirely due to the electrostatic interaction between small and 1D species as specified by the second term in Eq.~(\ref{PE1}). Given this, we cannot exclude that the ``inverted Marcus regime'' \cite{nitzan} could be taking place here over some range of energy offsets. To verify this conjecture, however, one would have to extend the current framework to explicitly include a multitude of electronic states on the 1D species, and we do not show it in the figure. The reverse $2 \rightarrow 1$ transition into a discrete electronic level of the small species features the inverted regime as usual \cite{nitzan}.

\section{Transfer between two 1D semiconductors}

The framework described in Sec.~\ref{SSLS} can be easily generalized to electron transfer between two parallel 1D semiconductors (intertube or interchain transfer). Realizing its likely narrow region of applicability in terms of the offset of equilibrium energies (limited by the restriction to the lowest-energy electron states only), we will provide here just a brief outline. In what follows index $i$ assumes values of 1 and 2 for the ``donor'' and ``acceptor'' systems. As the state of polarization is now described by two fictitious charge densities:
$z_i = \int dx\, \rfix$, the energy stored in polarization is given now  by
$$
U = \frac{1}{2}\sum_{ij}\int dx\, dy\, \rfix\, V_{ij} (x-y)\, \rfjy.
$$
Purely electronic states are derived from two Schr\"{o}dinger equations
\begin{equation}\label{TTSE}
\Hix\wfix = (E_i-\Ebr{i})\wfix
\end{equation}
with their respective Hamiltonians:
$$
\Hix = -\frac{\hbar^2}{2m_i}\secder + \PEi (x).
$$
Each is composed of the kinetic energy with the corresponding effective mass $m_i$ and the potential energy due to the solvent polarization:
$$
\PEi (x)=-q\sum_j \int dy\ V_{ij} (x-y)\, \rfjy.
$$
The lowest-energy solutions to Eqs.~(\ref{TTSE}) yield the real charge densities $\rhix = q \norm{\wfix}$.

Following the same example of tubular charge distributions (radii $R_i$), the electrostatic kernels would be described by the following Fourier-transforms:
$$
\Vf_{ii} (k) = 2I_0(kR_i)K_0(kR_i)/\est
$$
for the intratube interactions (the same as Eq.~(\ref{Kern1})), and
\begin{equation}\label{Kern4}
\Vf_{12} (k) = 2I_0(kR_1)I_0(kR_2)K_0(kd)/\est
\end{equation}
for the intertube interaction (instead of Eq.~(\ref{Kern2})), where $d$ is the distance between axes of the tubes.

As in the earlier discussion, the unconstrained (in electronic energy) minimization of total energies with the electron in state 1 or in state 2 results in the equilibrated polarons discussed in Sec.~\ref{Solv1} with some binding energies. The optimal electron transition state is determined under the constraint of equal electronic energies by minimizing the auxiliary functional
\begin{equation}\label{FU3}
\tEto = U + E_1 + \al (E_2 - E_1) + \sum_i \be_i \NN_i,
\end{equation}
where $\NN_i$ is the normalization condition (\ref{Norm}) for wavefunction $\wfix$. One readily finds that the minimum of Eq.~(\ref{FU3}) is achieved for the fictitious densities satisfying
$$
\rfox = (1-\al) \rhox, \ \ \ \ \rftx = \al \rhtx,
$$
that is, again obeying the partitioning (\ref{part1}) of the charge $q$ at the transition point. One convenient approach to solving the  problem of optimal fluctuations described by above equations is to specify the partitioning parameter $\al$ and then solve equations self-consistently to find the required energy offset and activation energies.

\begin{figure}
\centering
\includegraphics[scale=0.6]{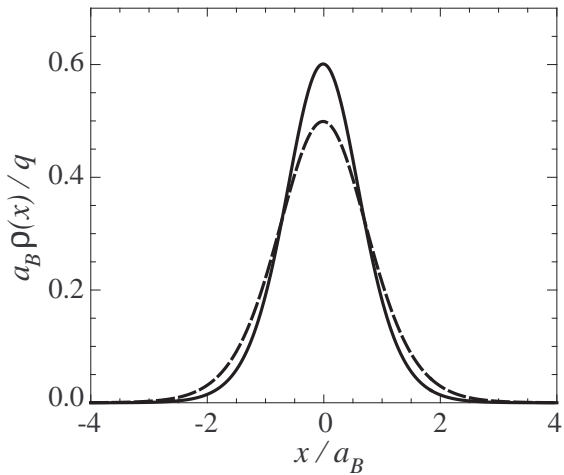}
\caption{\label{TTDens} Example of calculated electronic charge densities $\rhx$ in the equilibrated polaron (solid line) and at the electron transition point (dashed line) for transfer between two identical tubes. Parameters used here are $R=8$ \AA, $d=19.5$ \AA, and $\aB=32$ \AA.}
\end{figure}

A special interesting case is the transfer between two identical 1D species, to which we restrict our comments here. In this case the optimal transition fluctuation corresponds to $\al=1/2$. As an example of calculational results, Fig.~\ref{TTDens} compares electronic density in the equilibrated polaron and at the electron transition point between two identical tubes for a set of representative parameters. As one should expect on physical grounds, the electronic density spreads out at the transition. Once again, the magnitude of the effect here is ``mitigated'' by the fact of relatively appreciable intertube electrostatic interaction (\ref{Kern4}). From our calculations for a region of ``physically reasonable'' parameter values, we find that the activation energy in these cases is ordinarily a fraction (roughly 0.25--0.35) of the individual polaron binding energy $\Ebi$. If the latter is on the order of 0.1 eV, this assessment would indicate that activation energies are comparable to ambient temperatures making the thermally activated transfer between identical 1D species efficient.

\section{Discussion}

When a small-diameter 1D semiconductor nanostructure is immersed in a sluggish polar medium (solvent), fluctuations of the medium polarization may result in the appearance of localized electronic levels inside the band gap. An excess charge carrier (an electron or a hole) can occupy such a level and undergo self-localization into a large-radius adiabatic polaron accompanied by a self-consistent polarization pattern. Estimates based on representative parameter values indicate that the binding energy $\Ebi$ of solvation-induced 1D polarons is expected to be \textit{on the order of} 0.1 eV. Such estimates appear in agreement with our assessment \cite{YNGpol, polcylinder} that the polaron binding may reach a substantial fraction, roughly one-third, of the binding energy of well-known Wannier-Mott excitons in these structures, and experimental reports \cite{optrescnt,bachilo_bind,krauss1} of the latter in SWCNTs as 0.4--0.6 eV. The physical properties of strong-coupling polarons are quite different from the nearly free band electrons frequently discussed in the context of electronic transport in nanotubes and nanowires. One particularly notes that the mobility of solvated charge carriers is drastically reduced due to the dissipative drag of the medium \cite{basko,CBDrag,GU_lowfreq}. On the other hand, thermal dissociation of excitons into electron- and hole-polaron pairs may be enhanced \cite{YNGpol}.

In this paper we addressed effects of the solvent polarization on the fundamental process of electron transfer to/from 1D semiconductors within an appropriately adapted framework of the Marcus theory of thermally activated transfer. The resulting qualitative picture shows that, for a range of energy offsets, the transfer can indeed take place via solvation-induced polaronic-like states whose energies are below the band continuum, at variance with the picture of transfer into rigid-band states unaffected by the polarization of the medium. In this sense, the transfer we discussed is closer to the traditional scenario of transfer between two discrete electronic levels, both modulated by medium fluctuations. A qualitative difference with the fixed-shape electronic states of small species, however, is that the electronic density on 1D species is variable and can adjust its spatial extent during the transfer. The framework we employed here considers only lowest-energy electronic states, and it would have to be extended with account of a multitude of states in order to treat wider ranges of energy offsets and to verify if the inverted Marcus regime may be realizable.

Similarly to approach of Ref.~\cite{Basko_injection}, our analysis was based on looking for optimal fluctuations of the medium polarization that promote iso-energetic electron transfer. Knowing the (free) energy of such fluctuations allows one to assess the corresponding thermal activation energies $\Eact$. The optimal fluctuations correspond to the maximum of the integrand in Eq.~(\ref{FermiGR1}) and determine the exponential factor $\exp\left(-\Eact/\kB T \right)$ in the transfer rate, ordinarily of most interest in electron transfer problems \cite{nitzan}. The calculation of prefactors is more involved requiring considerations of polarization patterns ``in the vicinity'' of the optimal state at the transition. We refer the reader to Ref.~\cite{Basko_injection} for a discussion of some aspects of such calculations and just note here that the prefactors do not bring any ``extraordinary'' contributions \footnote{The only technical issue requiring care is a treatment of so-called zero-frequency translational modes appearing in the continuum polaron models (see, e.~g., references in Ref.~\cite{GU_lowfreq}) as a consequence of the translational invariance of the polaron center position. In the problem of electron injection at the end of a polymer chain discussed in Ref.~\cite{Basko_injection}, the authors took into account the fact that the translational invariance is broken at the end. In our case, one can make similar arguments regarding the position of the transitional charge density with respect to the given position of the small species or with respect to the center of the equilibrated polaron. It would, however, suffice to notice a simple fact that the electronic transition matrix element $\tab$ in Eq.~(\ref{FermiGR1}) falls off exponentially with the distance. Hence, any polarization configurations in the integrand of Eq.~(\ref{FermiGR1}) corresponding to electronic densities ``far away'' from each other (along the 1D system) would have exponentially vanishing contributions to the integral.}.

Illustrative calculations in this paper have been performed for tubular 1D geometries with possible applications to carbon nanotube systems. The generic character of the picture should however be evident as applied to other transverse charge distribution geometries, such as in small-diameter nanowires or conjugated polymers.

\bibliography{fluctuations}

\begin{thebibliography}{35}
\expandafter\ifx\csname natexlab\endcsname\relax\def\natexlab#1{#1}\fi
\expandafter\ifx\csname bibnamefont\endcsname\relax
  \def\bibnamefont#1{#1}\fi
\expandafter\ifx\csname bibfnamefont\endcsname\relax
  \def\bibfnamefont#1{#1}\fi
\expandafter\ifx\csname citenamefont\endcsname\relax
  \def\citenamefont#1{#1}\fi
\expandafter\ifx\csname url\endcsname\relax
  \def\url#1{\texttt{#1}}\fi
\expandafter\ifx\csname urlprefix\endcsname\relax\def\urlprefix{URL }\fi
\providecommand{\bibinfo}[2]{#2}
\providecommand{\eprint}[2][]{\url{#2}}

\bibitem[{\citenamefont{Kuznetsov}(1995)}]{CTbook}
\bibinfo{author}{\bibfnamefont{A.~M.} \bibnamefont{Kuznetsov}},
  \emph{\bibinfo{title}{Charge Transfer in Physics, Chemistry and Biology}}
  (\bibinfo{publisher}{Gordon and Breach}, \bibinfo{address}{Luxembourg},
  \bibinfo{year}{1995}).

\bibitem[{\citenamefont{Kuznetsov and Ulstrup}(1999)}]{CTbook1}
\bibinfo{author}{\bibfnamefont{A.~M.} \bibnamefont{Kuznetsov}}
  \bibnamefont{and} \bibinfo{author}{\bibfnamefont{J.}~\bibnamefont{Ulstrup}},
  \emph{\bibinfo{title}{Electron Transfer in Chemistry and Biology}}
  (\bibinfo{publisher}{Wiley}, \bibinfo{address}{Chichester, England},
  \bibinfo{year}{1999}).

\bibitem[{\citenamefont{May and K\"{u}hn}(2004)}]{CET2004}
\bibinfo{author}{\bibfnamefont{V.}~\bibnamefont{May}} \bibnamefont{and}
  \bibinfo{author}{\bibfnamefont{O.}~\bibnamefont{K\"{u}hn}},
  \emph{\bibinfo{title}{Charge and Energy Transfer Dynamics in Molecular
  Systems}} (\bibinfo{publisher}{Wiley-VCH}, \bibinfo{address}{Weinheim,
  Germany}, \bibinfo{year}{2004}).

\bibitem[{\citenamefont{Fawcett}(2004)}]{fawcett}
\bibinfo{author}{\bibfnamefont{W.~R.} \bibnamefont{Fawcett}},
  \emph{\bibinfo{title}{Liquids, solutions and interfaces}}
  (\bibinfo{publisher}{Oxford}, \bibinfo{address}{Oxford},
  \bibinfo{year}{2004}).

\bibitem[{\citenamefont{Nitzan}(2006)}]{nitzan}
\bibinfo{author}{\bibfnamefont{A.}~\bibnamefont{Nitzan}},
  \emph{\bibinfo{title}{Chemical Dynamics in Condensed Phases}}
  (\bibinfo{publisher}{Oxford}, \bibinfo{address}{New York},
  \bibinfo{year}{2006}).

\bibitem[{\citenamefont{Barford}(2005)}]{Barford_book}
\bibinfo{author}{\bibfnamefont{W.}~\bibnamefont{Barford}},
  \emph{\bibinfo{title}{Electronic and Optical Properties of Conjugated
  Polymers}} (\bibinfo{publisher}{Oxford University Press},
  \bibinfo{address}{Oxford}, \bibinfo{year}{2005}).

\bibitem[{\citenamefont{Licht}(2002)}]{SCelectrodes}
\bibinfo{editor}{\bibfnamefont{S.}~\bibnamefont{Licht}}, ed.,
  \emph{\bibinfo{title}{Semiconductor Electrodes and Photoelectrochemistry}}
  (\bibinfo{publisher}{Wiley-VCH}, \bibinfo{address}{Weinheim, Germany},
  \bibinfo{year}{2002}).

\bibitem[{\citenamefont{Landau and Lifshitz}(1981)}]{LL3}
\bibinfo{author}{\bibfnamefont{L.~D.} \bibnamefont{Landau}} \bibnamefont{and}
  \bibinfo{author}{\bibfnamefont{E.~M.} \bibnamefont{Lifshitz}},
  \emph{\bibinfo{title}{Quantum Mechanics. Non-relativistic Theory}}
  (\bibinfo{publisher}{Butterworth-Heinemann}, \bibinfo{address}{Oxford},
  \bibinfo{year}{1981}).

\bibitem[{\citenamefont{Basko and Conwell}(2002{\natexlab{a}})}]{basko}
\bibinfo{author}{\bibfnamefont{D.~M.} \bibnamefont{Basko}} \bibnamefont{and}
  \bibinfo{author}{\bibfnamefont{E.~M.} \bibnamefont{Conwell}},
  \bibinfo{journal}{Phys. Rev. Lett.} \textbf{\bibinfo{volume}{88}},
  \bibinfo{pages}{098102} (\bibinfo{year}{2002}{\natexlab{a}}).

\bibitem[{\citenamefont{Gartstein}(2006)}]{YNGpol}
\bibinfo{author}{\bibfnamefont{Y.~N.} \bibnamefont{Gartstein}},
  \bibinfo{journal}{Phys. Lett. A} \textbf{\bibinfo{volume}{349}},
  \bibinfo{pages}{377} (\bibinfo{year}{2006}).

\bibitem[{\citenamefont{Gartstein et~al.}(2007)\citenamefont{Gartstein,
  Bustamante, and {Ortega Castillo}}}]{polcylinder}
\bibinfo{author}{\bibfnamefont{Y.~N.} \bibnamefont{Gartstein}},
  \bibinfo{author}{\bibfnamefont{T.~D.} \bibnamefont{Bustamante}},
  \bibnamefont{and} \bibinfo{author}{\bibfnamefont{S.}~\bibnamefont{{Ortega
  Castillo}}}, \bibinfo{journal}{J. Phys.: Condens. Matter}
  \textbf{\bibinfo{volume}{19}}, \bibinfo{pages}{156210}
  (\bibinfo{year}{2007}).

\bibitem[{\citenamefont{Gartstein and Ussery}(2008)}]{GU_lowfreq}
\bibinfo{author}{\bibfnamefont{Y.~N.} \bibnamefont{Gartstein}}
  \bibnamefont{and} \bibinfo{author}{\bibfnamefont{G.~L.}
  \bibnamefont{Ussery}}, \bibinfo{journal}{Phys. Lett. A}
  \textbf{\bibinfo{volume}{372}}, \bibinfo{pages}{5909} (\bibinfo{year}{2008}).

\bibitem[{\citenamefont{Ussery and Gartstein}(2009)}]{UG_optabs}
\bibinfo{author}{\bibfnamefont{G.~L.} \bibnamefont{Ussery}} \bibnamefont{and}
  \bibinfo{author}{\bibfnamefont{Y.~N.} \bibnamefont{Gartstein}},
  \bibinfo{journal}{J. Chem. Phys.} \textbf{\bibinfo{volume}{130}},
  \bibinfo{pages}{014701} (\bibinfo{year}{2009}).

\bibitem[{\citenamefont{Mayo and Gartstein}(2008)}]{MG_abinitio1}
\bibinfo{author}{\bibfnamefont{M.~L.} \bibnamefont{Mayo}} \bibnamefont{and}
  \bibinfo{author}{\bibfnamefont{Y.~N.} \bibnamefont{Gartstein}},
  \bibinfo{journal}{Phys. Rev. B} \textbf{\bibinfo{volume}{78}},
  \bibinfo{pages}{073402} (\bibinfo{year}{2008}).

\bibitem[{\citenamefont{Mayo and Gartstein}(2009)}]{MG_abinitio2}
\bibinfo{author}{\bibfnamefont{M.~L.} \bibnamefont{Mayo}} \bibnamefont{and}
  \bibinfo{author}{\bibfnamefont{Y.~N.} \bibnamefont{Gartstein}},
  \bibinfo{journal}{J. Chem. Phys.} \textbf{\bibinfo{volume}{130}},
  \bibinfo{pages}{134705} (\bibinfo{year}{2009}).

\bibitem[{\citenamefont{Law et~al.}(2005)\citenamefont{Law, Greene, Johnson,
  Saykally, and Yang}}]{pyang05}
\bibinfo{author}{\bibfnamefont{M.}~\bibnamefont{Law}},
  \bibinfo{author}{\bibfnamefont{L.~E.} \bibnamefont{Greene}},
  \bibinfo{author}{\bibfnamefont{J.~C.} \bibnamefont{Johnson}},
  \bibinfo{author}{\bibfnamefont{R.}~\bibnamefont{Saykally}}, \bibnamefont{and}
  \bibinfo{author}{\bibfnamefont{P.}~\bibnamefont{Yang}},
  \bibinfo{journal}{Nature Materials} \textbf{\bibinfo{volume}{4}},
  \bibinfo{pages}{455} (\bibinfo{year}{2005}).

\bibitem[{\citenamefont{Paulose et~al.}(2006)\citenamefont{Paulose, Mor,
  Varghese, Shankar, and Grimes}}]{photcat2}
\bibinfo{author}{\bibfnamefont{M.}~\bibnamefont{Paulose}},
  \bibinfo{author}{\bibfnamefont{G.}~\bibnamefont{Mor}},
  \bibinfo{author}{\bibfnamefont{O.}~\bibnamefont{Varghese}},
  \bibinfo{author}{\bibfnamefont{K.}~\bibnamefont{Shankar}}, \bibnamefont{and}
  \bibinfo{author}{\bibfnamefont{C.}~\bibnamefont{Grimes}},
  \bibinfo{journal}{J. Photochem. Photobiol., A}
  \textbf{\bibinfo{volume}{178}}, \bibinfo{pages}{8} (\bibinfo{year}{2006}).

\bibitem[{\citenamefont{Kamat}(2007)}]{kamat07}
\bibinfo{author}{\bibfnamefont{P.~V.} \bibnamefont{Kamat}},
  \bibinfo{journal}{J. Phys. Chem. C} \textbf{\bibinfo{volume}{111}},
  \bibinfo{pages}{2834} (\bibinfo{year}{2007}).

\bibitem[{\citenamefont{Liu et~al.}(2008)\citenamefont{Liu, Cao, Yang, Wang,
  Dubois, Zhou, Graff, Pederson, and Zhang}}]{csc08}
\bibinfo{author}{\bibfnamefont{J.}~\bibnamefont{Liu}},
  \bibinfo{author}{\bibfnamefont{G.}~\bibnamefont{Cao}},
  \bibinfo{author}{\bibfnamefont{Z.}~\bibnamefont{Yang}},
  \bibinfo{author}{\bibfnamefont{D.}~\bibnamefont{Wang}},
  \bibinfo{author}{\bibfnamefont{D.}~\bibnamefont{Dubois}},
  \bibinfo{author}{\bibfnamefont{X.}~\bibnamefont{Zhou}},
  \bibinfo{author}{\bibfnamefont{G.}~\bibnamefont{Graff}},
  \bibinfo{author}{\bibfnamefont{L.}~\bibnamefont{Pederson}}, \bibnamefont{and}
  \bibinfo{author}{\bibfnamefont{J.}~\bibnamefont{Zhang}},
  \bibinfo{journal}{ChemSusChem} \textbf{\bibinfo{volume}{1}},
  \bibinfo{pages}{676} (\bibinfo{year}{2008}).

\bibitem[{\citenamefont{Wallace et~al.}(2009)\citenamefont{Wallace, Chen,
  Mozer, Forsyth, MacFarlane, and Wang}}]{mattod09}
\bibinfo{author}{\bibfnamefont{G.}~\bibnamefont{Wallace}},
  \bibinfo{author}{\bibfnamefont{J.}~\bibnamefont{Chen}},
  \bibinfo{author}{\bibfnamefont{A.}~\bibnamefont{Mozer}},
  \bibinfo{author}{\bibfnamefont{M.}~\bibnamefont{Forsyth}},
  \bibinfo{author}{\bibfnamefont{D.}~\bibnamefont{MacFarlane}},
  \bibnamefont{and} \bibinfo{author}{\bibfnamefont{C.}~\bibnamefont{Wang}},
  \bibinfo{journal}{Mater. Today} \textbf{\bibinfo{volume}{12}},
  \bibinfo{pages}{No.~6, 20} (\bibinfo{year}{2009}).

\bibitem[{\citenamefont{Yu and Chen}(2009)}]{yurev}
\bibinfo{author}{\bibfnamefont{K.}~\bibnamefont{Yu}} \bibnamefont{and}
  \bibinfo{author}{\bibfnamefont{J.}~\bibnamefont{Chen}},
  \bibinfo{journal}{Nanoscale Res. Lett.} \textbf{\bibinfo{volume}{4}},
  \bibinfo{pages}{1} (\bibinfo{year}{2009}).

\bibitem[{\citenamefont{Vichchulada et~al.}(2009)\citenamefont{Vichchulada,
  Lipscomb, Zhang, and Lay}}]{sens1}
\bibinfo{author}{\bibfnamefont{P.}~\bibnamefont{Vichchulada}},
  \bibinfo{author}{\bibfnamefont{L.}~\bibnamefont{Lipscomb}},
  \bibinfo{author}{\bibfnamefont{Q.}~\bibnamefont{Zhang}}, \bibnamefont{and}
  \bibinfo{author}{\bibfnamefont{M.}~\bibnamefont{Lay}}, \bibinfo{journal}{J.
  Nanosci. Nanotech.} \textbf{\bibinfo{volume}{9}}, \bibinfo{pages}{2189}
  (\bibinfo{year}{2009}).

\bibitem[{\citenamefont{O'Connell et~al.}(2005)\citenamefont{O'Connell,
  Eibergen, and Doorn}}]{chirsel}
\bibinfo{author}{\bibfnamefont{M.~J.} \bibnamefont{O'Connell}},
  \bibinfo{author}{\bibfnamefont{E.~E.} \bibnamefont{Eibergen}},
  \bibnamefont{and} \bibinfo{author}{\bibfnamefont{S.~K.} \bibnamefont{Doorn}},
  \bibinfo{journal}{Nature Materials} \textbf{\bibinfo{volume}{4}},
  \bibinfo{pages}{412} (\bibinfo{year}{2005}).

\bibitem[{\citenamefont{Choi and Strano}(2007)}]{CNTsolvchrom}
\bibinfo{author}{\bibfnamefont{J.~H.} \bibnamefont{Choi}} \bibnamefont{and}
  \bibinfo{author}{\bibfnamefont{M.~S.} \bibnamefont{Strano}},
  \bibinfo{journal}{Appl. Phys. Lett.} \textbf{\bibinfo{volume}{90}},
  \bibinfo{pages}{223114} (\bibinfo{year}{2007}).

\bibitem[{\citenamefont{Basko and
  Conwell}(2002{\natexlab{b}})}]{Basko_injection}
\bibinfo{author}{\bibfnamefont{D.~M.} \bibnamefont{Basko}} \bibnamefont{and}
  \bibinfo{author}{\bibfnamefont{E.~M.} \bibnamefont{Conwell}},
  \bibinfo{journal}{Phys. Rev. B} \textbf{\bibinfo{volume}{66}},
  \bibinfo{pages}{094304} (\bibinfo{year}{2002}{\natexlab{b}}).

\bibitem[{\citenamefont{Pennington and Goldsman}(2005)}]{NTEffectMass}
\bibinfo{author}{\bibfnamefont{G.}~\bibnamefont{Pennington}} \bibnamefont{and}
  \bibinfo{author}{\bibfnamefont{N.}~\bibnamefont{Goldsman}},
  \bibinfo{journal}{Phys. Rev. B} \textbf{\bibinfo{volume}{71}},
  \bibinfo{pages}{205318} (\bibinfo{year}{2005}).

\bibitem[{\citenamefont{Pedersen}(2003)}]{pedersen}
\bibinfo{author}{\bibfnamefont{T.~G.} \bibnamefont{Pedersen}},
  \bibinfo{journal}{Phys. Rev. B} \textbf{\bibinfo{volume}{67}},
  \bibinfo{pages}{073401} (\bibinfo{year}{2003}).

\bibitem[{\citenamefont{Perebeinos et~al.}(2004)\citenamefont{Perebeinos,
  Tersoff, and Avouris}}]{tersscaling}
\bibinfo{author}{\bibfnamefont{V.}~\bibnamefont{Perebeinos}},
  \bibinfo{author}{\bibfnamefont{J.}~\bibnamefont{Tersoff}}, \bibnamefont{and}
  \bibinfo{author}{\bibfnamefont{P.}~\bibnamefont{Avouris}},
  \bibinfo{journal}{Phys. Rev. Lett.} \textbf{\bibinfo{volume}{92}},
  \bibinfo{pages}{257402} (\bibinfo{year}{2004}).

\bibitem[{\citenamefont{Haug and Koch}(2004)}]{haugbook}
\bibinfo{author}{\bibfnamefont{H.}~\bibnamefont{Haug}} \bibnamefont{and}
  \bibinfo{author}{\bibfnamefont{S.~W.} \bibnamefont{Koch}},
  \emph{\bibinfo{title}{Quantum theory of the optical and electronic properties
  of semiconductors}} (\bibinfo{publisher}{World Scientific},
  \bibinfo{address}{New Jersey}, \bibinfo{year}{2004}).

\bibitem[{\citenamefont{Ando}(1997)}]{ando}
\bibinfo{author}{\bibfnamefont{T.}~\bibnamefont{Ando}}, \bibinfo{journal}{J.
  Phys. Soc. Japan} \textbf{\bibinfo{volume}{66}}, \bibinfo{pages}{1066}
  (\bibinfo{year}{1997}).

\bibitem[{\citenamefont{Jackson}(1998)}]{jackson}
\bibinfo{author}{\bibfnamefont{J.~D.} \bibnamefont{Jackson}},
  \emph{\bibinfo{title}{Classical electrodynamics}}
  (\bibinfo{publisher}{Wiley}, \bibinfo{address}{New York},
  \bibinfo{year}{1998}).

\bibitem[{\citenamefont{Wang et~al.}(2005)\citenamefont{Wang, Dukovic, Brus,
  and Heinz}}]{optrescnt}
\bibinfo{author}{\bibfnamefont{F.}~\bibnamefont{Wang}},
  \bibinfo{author}{\bibfnamefont{G.}~\bibnamefont{Dukovic}},
  \bibinfo{author}{\bibfnamefont{L.~E.} \bibnamefont{Brus}}, \bibnamefont{and}
  \bibinfo{author}{\bibfnamefont{T.~F.} \bibnamefont{Heinz}},
  \bibinfo{journal}{Science} \textbf{\bibinfo{volume}{308}},
  \bibinfo{pages}{838} (\bibinfo{year}{2005}).

\bibitem[{\citenamefont{Ma et~al.}(2005)\citenamefont{Ma, Valkunas, Bachilo,
  and Fleming}}]{bachilo_bind}
\bibinfo{author}{\bibfnamefont{Y.-Z.} \bibnamefont{Ma}},
  \bibinfo{author}{\bibfnamefont{L.}~\bibnamefont{Valkunas}},
  \bibinfo{author}{\bibfnamefont{S.~M.} \bibnamefont{Bachilo}},
  \bibnamefont{and} \bibinfo{author}{\bibfnamefont{G.~R.}
  \bibnamefont{Fleming}}, \bibinfo{journal}{J. Phys. Chem. B}
  \textbf{\bibinfo{volume}{109}}, \bibinfo{pages}{15671}
  (\bibinfo{year}{2005}).

\bibitem[{\citenamefont{Wang et~al.}(2006)\citenamefont{Wang, Pedrosa, Krauss,
  and Rothberg}}]{krauss1}
\bibinfo{author}{\bibfnamefont{Z.}~\bibnamefont{Wang}},
  \bibinfo{author}{\bibfnamefont{H.}~\bibnamefont{Pedrosa}},
  \bibinfo{author}{\bibfnamefont{T.}~\bibnamefont{Krauss}}, \bibnamefont{and}
  \bibinfo{author}{\bibfnamefont{L.}~\bibnamefont{Rothberg}},
  \bibinfo{journal}{Phys. Rev. Lett.} \textbf{\bibinfo{volume}{96}},
  \bibinfo{pages}{047403} (\bibinfo{year}{2006}).

\bibitem[{\citenamefont{Conwell and Basko}(2006)}]{CBDrag}
\bibinfo{author}{\bibfnamefont{E.~M.} \bibnamefont{Conwell}} \bibnamefont{and}
  \bibinfo{author}{\bibfnamefont{D.~M.} \bibnamefont{Basko}},
  \bibinfo{journal}{J. Phys. Chem. B} \textbf{\bibinfo{volume}{110}},
  \bibinfo{pages}{23603} (\bibinfo{year}{2006}).

\end{thebibliography}

\end{document}